\documentclass{elsart}

\begin{document}

\journal{Physics Letters A}
\date{4 February 1997}

\begin{frontmatter}
\title{Comment on\\
``Self-organized criticality in living systems''\\
by C. Adami}
\author[SFI]{M. E. J. Newman},
\author[SFI]{Simon M. Fraser},
\author[SFI,Nordita]{Kim Sneppen}\\
and
\author[SFI]{William A. Tozier}
\address[SFI]{Santa Fe Institute, 1399 Hyde Park Road, Santa Fe, NM 87501.
U.S.A.}
\address[Nordita]{Nordita, Blegdamsvej 17, DK-2100 Copenhagen \O.
Denmark.}
\begin{abstract}
  Following extensive numerical experiments, it has been suggested
  that the evolution of competing computer programs in artificial life
  simulations shows signs of being a self-organized critical process.
  The primary evidence for this claim comes from the distribution of
  the lifetimes of species in the simulations, which appears to follow
  a power law.  We argue that, for a number of reasons, it is unlikely
  that the system is in fact at a critical point and suggest an
  alternative explanation for the power-law lifetime distribution.
\end{abstract}
\end{frontmatter}

In two recent papers, Adami~\cite{Adami95a,Adami95b} has presented
results from extensive simulations using the Tierra artificial life system,
which appear to show evidence of self-organized critical behaviour in an
evolving system.  Should it turn out to be correct, this observation would
be of considerable interest in evolutionary theory, where the idea of
self-organization has been a topic of debate for some years now.  In this
comment, however, we would like to question this interpretation of the
Tierra data.  For a number of reasons, we believe it unlikely that one
would be able to see critical phenomena in a system such as Tierra, and
furthermore there are certain features of the data which strongly suggest
that the system is not in a critical state.  We therefore propose an
alternative interpretation which, we believe, explains the observed
behaviour of the system without assuming criticality.

The fundamental idea behind self-organized critical models of evolving
systems is that of coevolutionary avalanches~\cite{Kauffman92}.  It
is known that species interaction through such mechanisms as competition,
predation, parasitism and symbiosis can cause one species to evolve as a
result of the evolution of another.  It is possible therefore for the
chance mutation of a species in a large ecosystem to spark off a chain, or
``avalanche'', of evolution affecting a large number of other species and
potentially the entire ecosystem.  Models of this process have been
proposed, notably those of Kauffman and Johnsen~\cite{KJ91} and of
Bak and Sneppen~\cite{BS93}, which predict that the resulting
reorganizations of the ecosystem will drive it towards a critical state in
which the distribution of avalanche sizes follows a power law.

Coevolutionary avalanches have not been directly observed in Tierra.  The
primary evidence for criticality comes instead from the distribution of
``epoch'' lengths in the evolution of the ecosystem.  If we record the
fitness (which can be defined roughly as the reproduction rate) of the
fittest species in Tierra as a function of time, the measured values
increase not smoothly, but in jumps, which arise as the result of
evolutionary innovations in the population.  Adami~\cite{Adami95a}
has dubbed these jumps ``phase transitions'', by analogy with the first
order phase transitions of statistical physics.  Immediately before an
innovation the system can be considered to be in a metastable state,
waiting for the fluctuation which will take it over some barrier to a state
of lower free energy.  The transition itself proceeds precisely by a
nucleation process, just as do many more familiar transitions.  One
individual in the population discovers the trick which allows it to
reproduce more efficiently and, like the freezing of super-cooled water,
the offspring of that individual spread through the ecosystem causing a
sudden jump in the mean fitness of the population.  (Notice though that in
Tierra the jump is merely to another metastable state which will ultimately
itself give way to a still more favourable one.)  If one constructs a
histogram of the time intervals $\tau$, also called epochs, between each
jump and the next, the resulting curve follows a power-law distribution
whose exponent is measured to be
$\alpha=1.10\pm0.05$~\cite{Adami95a}.  Power-law distributions are
often taken to be indicative of self-organized critical behaviour, and
indeed the evolution model of Bak and Sneppen~\cite{BS93} predicts a
power law of precisely this type in the lengths of the epochs separating
consecutive avalanches.  Do the data therefore indicate that the Tierra
system evolves to a self-organized critical state?  We have a number of
objections to this interpretation.

(i) The jumps, or phase transitions, seen in Tierra are not equivalent to
coevolutionary avalanches.  Whilst avalanches arise through the interaction
of many species, the phase transitions here are essentially the product of
just one species, whose fitness exceeds that of all others, allowing it
rapidly to dominate the system.

(ii) The phase transitions are strongly first order, as is clear from the
presence of fast nucleation processes.  Critical phenomena are not observed
in the vicinity of strongly first order transitions.

(iii) In our own investigations of the Tierra system, we observe that
inter-species interactions are relatively rare.  On average each species
interacts with less than one other.  However, each species must interact
with at least two others in order to establish a percolating interaction
network.  Such a network is necessary if system-spanning coevolutionary
avalanches are to take place, and therefore we conclude that avalanches of
this kind are not possible in Tierra.  In real ecosystems it is estimated
that each species interacts with between three and four others on the
average~\cite{Sugihara89}, a figure more consistent with the
coevolutionary avalanche picture.  (We can neglect the trivial interaction
that arises because all species in Tierra compete for space and CPU
time---this interaction produces correlations between the birth and death
rates of different species, but does not give rise to coevolution.)

(iv) The Tierran ecology at any one moment tends to be dominated by the
fittest species, or the fittest handful of species.  Even ignoring points
(i), (ii) and (iii), this would limit the size of possible avalanches to
just a few species.  In effect, the finite-size effects on the system would
truncate any critical behaviour so that we would not expect it to be
visible in the simulations.

So, if the Tierra system is not in a critical state, how are we to account
for the appearance of the power law?  One possible explanation is that
Tierran evolution is an ``extremal random process''.  Suppose that at some
point during a simulation the process of finding a genotype which is fitter
than the current fittest one in the system takes a time $t_1$.  How long
then will it take to find another one which is fitter still?  If the
sampling of fitness by mutation and selection is a random process, we can
assume that it will on average take another time $t_1$ to find a new
genotype with the same or greater fitness, or an aggregate time of $t_2 =
2t_1$.  This result will be true regardless of the distribution over
fitnesses which we sample---for instance, it makes no difference if we
sample genotypes of lower fitness more often than those of higher fitness,
the result will still be the same.  Iterating the argument, the time taken
to find the next fitter genotype will be $t_3 = 2t_2 = 4t_1$ and so forth.
The time $t_n$ at which the $n^{\rm th}$ such innovation occurs is then
given by
\begin{equation}
t_n = t_1 2^n,
\label{times}
\end{equation}
and the duration of the corresponding epoch is
\begin{equation}
\tau = t_{n+1} - t_n = t_1 (2^{n+1} - 2^n) = t_1 2^n.
\label{crucial}
\end{equation}
Thus the number of epoch lengths ${\rm d}n$ falling, on average, in an
interval ${\rm d}\tau$ is
\begin{equation}
{\rm d}n = p(\tau) {\rm d}\tau = {2{\rm d}\tau\over n\tau}.
\label{powerlaw}
\end{equation}
Equation~(\ref{crucial}) tells us that the $n$ in the denominator here goes
like $\log\tau$ and hence, apart from logarithmic corrections, the
distribution of epochs $p(\tau)$ goes like $\tau^{-1}$.  As we mentioned,
the actual measured behaviour is $\tau^{-\alpha}$ with
$\alpha=1.10\pm0.05$, so this result is in reasonable agreement with the
simulations.

An alternative explanation is that the Tierra system contains ``fitness
barriers'' across which species have to pass by mutation in order to reach
genotypes of higher fitness.  Models such as that of Bak and
Sneppen~\cite{BS93} assume that barrier crossing is a thermal
excitation process with a typical crossing time $\tau$ (equivalent to the
epoch lengths above) which is exponentially related to the barrier height
$B$:
\begin{equation}
\tau \propto {\rm e}^{-B/T},
\end{equation}
where $T$ is some temperature-like parameter whose exact value we do not
know.  If we assume that the barriers are randomly distributed according to
some probability distribution $p_{\rm barrier}(B)$, then the distribution
of $\tau$ is given by
\begin{equation}
p(\tau) = p_{\rm barrier}(B) {{\rm d}B\over{\rm d}\tau} \sim
          {p_{\rm barrier}(\log\tau)\over\tau}.
\end{equation}
So again, apart from logarithmic corrections, we find a distribution of
epoch lengths which goes like $\tau^{-1}$, in agreement with the data.

Either of these two scenarios is capable of explaining the behaviour
observed in Tierra without assuming self-organized criticality.  It would
be more satisfying however if we were able to say with some degree of
certainty which one we believe to be correct.  In the first scenario we can
use Equation~(\ref{times}) to predict the distribution of the jumps as a
function of the elapsed time during the simulation.  Following precisely
the same argument as we employed to write Equation~(\ref{powerlaw}) we can
then show that, if the first scenario is correct, the timing of the jumps
should, like the epoch lengths, follow a power law of $t^{-1}$.
Conversely, the distribution of jumps as a function of elapsed time in the
second scenario is entirely uniform.  Appropriate analysis of the data from
the simulations performed by Adami should therefore allow us to distinguish
between the two cases.

To conclude, we believe for a number of reasons that it is unlikely that
one can observe critical behaviour in a system such as Tierra.  The
observed power-law distribution of epoch times we explain instead as a
result of the random sampling of genotypes by the mutation and selection
processes, or possibly as the result of the way in which the system
traverses fitness barriers.  There is, nonetheless, much of interest in the
simulations carried out by Adami.  In particular, we believe that the
sudden jumps seen in the fitness constitute one of the clearest examples
yet of the phenomenon known as ``punctuated
equilibrium''~\cite{EG72}.  It is well known amongst palaeontologists
that, rather than evolving continually in small phenotypic jumps, fossil
species tend to remain unchanged for many millions of years, before
suddenly undergoing evolution to a new form.  An explanation of this effect
was given by George Simpson in 1944~\cite{Simpson44} who pointed out
that behaviour of exactly this type is to be expected of species whose
evolution to a fitter form requires them to cross some fitness barrier,
either energetic or entropic in nature.  Tierran organisms fall into
precisely this category, so it is satisfying to see punctuated behaviour so
clearly in evidence.

\end{document}